# On the discovery of meteoritic mineral Zolenskyite; The artificial origin should not be overlooked

B. P. Embaid

Laboratorio de Magnetismo, Escuela de Física, Universidad Central de Venezuela, Apartado 47586, Los Chaguaramos, Caracas 1041-A, Venezuela.

**Abstract**

Recently a new meteoritic mineral, Zolenskyite ($Fe_{0.99}Mn_{0.04}Ca_{0.01}Cr_{1.99}S_{3.98}$), was discovered from the Indarch meteorite. Zolenskyite was structurally indexed as the monoclinic C2/m $CrNb_2Se_4$ - $Cr_3S_4$ type structure of synthetic $FeCr_2S_4$, with unit cell parameters a = 12.84(1) Å, b = 3.44(1) Å, c = 5.94(1) Å and β = 117(1)°. Zolenskyite was reported as high-pressure phase formed from Daubréelite at high pressures and temperatures in highly shocked regions of the EH parent asteroid. Although this discovery provides valuable information about the origin of meteoritic mineral assemblages, the results and conclusions raise controversies with those reported in previous articles where the synthetic $FeCr_2S_4$ was described.

In this review, an alternative analysis of the supplementary X-ray data from Zolenskyite was made and yields to the monoclinic I2/m $Cr_3S_4$ type structure of synthetic $FeCr_2S_4$, with unit cell parameters a = 5.940 Å, b = 3.440 Å, c = 11.441 Å and β = 90.55°, in agreement with previous results in the literature and differ from those reported above.

Regarding the genesis of Zolenskyite and according to solid-state laboratory synthesis, the transformation of cubic $FeCr_2S_4$ phase (ideal composition of Daubréelite) to monoclinic $FeCr_2S_4$ (ideal composition of Zolenskyite) requires a process whose replication in Shock metamorphism stages is not well established and remains an open issue to address, together with another open issue related to the real composition of meteoritic minerals, with minor and trace metals, which is often overlooked when compared with synthetic ideal compositions. Whatever the results to be obtained by addressing the above open issues, they will shed more light on the genesis of Zolenskyite. In this context, it is worth to promote an open debate about the hypothesis of artificial origin.

## 1. INTRODUCTION

Ma and Rubin (2022) reported the discovery of Zolenskyite, ($Fe_{0.99}Mn_{0.04}Ca_{0.01}Cr_{1.99}S_{3.98}$), a new sulfide mineral from the Indarch meteorite. Zolenskyite was structurally indexed as the monoclinic C2/m $CrNb_2Se_4$ - $Cr_3S_4$ type structure of synthetic $FeCr_2S_4$, with unit cell parameters a = 12.84(1) Å, b = 3.44(1) Å, c = 5.94(1) Å and β = 117(1)°. According to the authors, Zolenskyite is a monoclinic polymorph of Daubréelite and may be a high-pressure phase, formed from Daubréelite at high pressures (several gigapascals) and moderate temperatures in highly shocked regions of the EH parent asteroid before becoming incorporated into Indarch via impact mixing. These results and conclusions raise controversies with those reported in a contemporary review article (Embaid 2022 and references therein) about the meteoritic minerals Heideite (Fe, Cr)$_{1+x}$ (Ti, Fe)$_2S_4$ from the Bustee and Kaidun meteorites and Brezinaite ($Cr_{2.65}Fe_{0.20}V_{0.09}Ti_{0.06}Mn_{0.04}$)$_{3.04}S_4$ from the Tucson meteorite, concerning their Solid State properties and genesis, where the synthetic $FeCr_2S_4$ was described.

The purpose of this review is to highlight the after mentioned controversies and disclose open issues to be addressed regarding the real or natural composition of meteoritic minerals, with minor and trace metals, which is usually overlooked when compared with synthetic ideal compositions, and hence, to seek for a unified description of Zolenskyite from the perspective of Solid State Physics and Planetary Science.

The following section of the manuscript (section 2) is focused on some solid state properties (crystallography and synthesis) of selected transition metal sulfides which represent the ideal composition of the meteoritic minerals Zolenskyite ($FeCr_2S_4$), Heideite ($FeTi_2S_4$) and Brezinaite ($Cr_3S_4$), this section can be considered as reference for the next sections 3 and 4 focused on the controversies and open issues to be addressed. The final section 5 is focused on the implications of the results to be obtained by addressing the above open issues, and aimed to promote an open debate about the hypothesis of artificial origin of the meteoritic mineral Zolenskyite, in the mainstream of the meteoritic minerals Heideite and Brezinaite, postulated as iron-based superconductors and extraterrestrial technosignatures.

## 2. CRYSTALLOGRAPHY AND SYNTHESIS OF $Cr_3S_4$, $FeCr_2S_4$ AND RELATED SULFIDES

### 2.1. $Cr_3S_4$, $FeTi_2S_4$ and $FeV_2S_4$

There are synthetic sulfide phases such as $FeV_2S_4$ (Embaid and Gonzalez-Jimenez 2013) and $FeTi_2S_4$ (Embaid et al. 2019) that are isostructural with the monoclinic $Cr_3S_4$ type structure (Space group I2/m, No 12) as shown in Fig. 1. This structure is metal-deficient NiAs-like, in which there are two crystallographic sites for cations, one site in a metal-deficient layer ($M_I$) with ordered metal vacancies and the second site in a metal full layer ($M_{II}$). These layers are intercalated between hexagonal close-packed sulfur layers. The lattice parameters of $Cr_3S_4$ are a = 5.964 Å, b = 3.428 Å, c = 11.272 Å and β

= 91.50 ° (Jellinek 1957). Regarding the synthetic sulfide phases $FeV_2S_4$ and $FeTi_2S_4$, Fe atoms are located in $M_I$ layer while V and Ti atoms are located in $M_{II}$ layer.

All these sulfides can be synthesized by annealing procedure (heating and slow cooling) (Jellinek 1957, Plovnick et al. 1968, Muranaka 1973, Oka et al. 1977, Bensch et al. 1986, Embaid and Gonzalez-Jimenez 2013, Embaid et al. 2019).

### 2.2. $FeCr_2S_4$

On the other side, the sulfide phase $FeCr_2S_4$, can be synthesized by a similar annealing procedure as mentioned above, but it crystallizes in the cubic spinel structure (space group F d -3 m S, No 227) (Bouchard et al. 1965, Greenwood and Whitfield 1968, El Goresy and Kullerud 1969, Göbel et al. 1975, Chen et al. 1999, Amiel et al. 2011). This phase undergoes a structural transformation to monoclinic $Cr_3S_4$ type structure if treated with temperature and pressure, then followed by quenching at room temperature; Bouchard (1967) reported the spinel – monoclinic transformation at 1000 °C and 6.5 GPa for one hour, Tressler et al. (1968) reported the same transformation at 520 °C and 5.5 GPa for seven days.

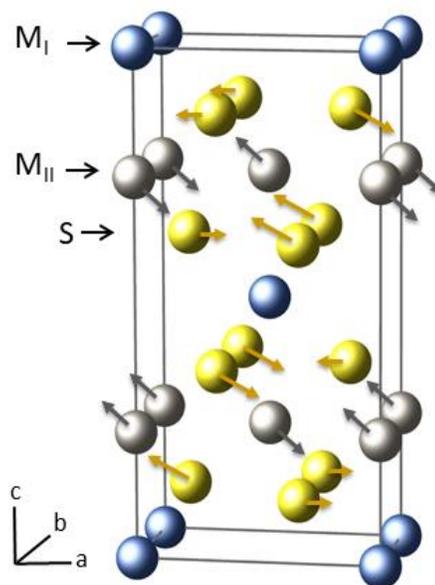

Fig. 1: The ideal structure of $Cr_3S_4$ (Space group I2/m, No. 12) after Jellinek (1957), blue spheres: metals in the metal deficient layer ($M_I$), grey spheres: metals in the metal full layer ($M_{II}$) and yellow spheres: sulfur atoms. The arrows indicate directions and magnitudes (five times enlarged) of the observed deviations from the ideal atomic positions, see text.

## 3. CONTROVERSIES TO ADDRESS REGARDING ZOLENSKYITE

### 3.1. Structural characterization

The d-spacing values included in the supplementary X-ray data from Zolenskyite provided by Ma and Rubin (2022) were tested with two software packages for indexing crystallographic structure; Treor90 (Werner et al. 1985) and Dicvol04 (Boultif and Louër 2004). Both tests led to identical results and point to a monoclinic lattice with multiple solutions having different cell parameters and the same minimum unit cell volume as shown in Table 1.

Table 1: Solutions for unit cell parameters of Zolenskyite

| Solution type | a (Å) | b (Å) | c (Å) | β (°) | Volume (Å$^3$) |
|---|---|---|---|---|---|
| Input cell * | 12.942 | 3.440 | 5.940 | 117.87 | 233.77 |
| Conventional cell ** | 11.441 | 3.440 | 5.940 | 90.55 | 233.77 |
| Reduced cell | 5.940 | 3.440 | 11.441 | 90.55 | 233.77 |

* Provided by Treor90 software.
** Also defined as direct cell parameters in Dicvol04 software.

The advantage of using Treor90 and Dicvol04 programs relies on the determination of the primitive cell, that is, the unit cell with the three shortest cell parameters that are the building blocks of the crystal. The first solution, called the input cell, with cell parameters similar to those reported by Ma and Rubin (2022) does not qualify as primitive cell because of larger cell parameters (a) and (β) if compared with those of the remaining solutions. The second and third solutions denoted as conventional cell and reduced cell are candidates for the primitive cell, and they differ by exchanging the values of cell parameters (a) and (c), which is characteristic of monoclinic system, as explained in the following:

In the monoclinic system, the inter-planar d-spacing ($d_{hkl}$) is given by the relation

$$\frac{1}{d_{hkl}^2} = \frac{1}{\sin^2 \beta}\left(\frac{h^2}{a^2} + \frac{k^2 \sin^2 \beta}{b^2} + \frac{l^2}{c^2} - \frac{2hl\cos\beta}{ac}\right) \quad (1)$$

It can be noted from Eq. (1) that $d_{hkl}$ is invariant by exchanging the values of (a) with (c) and the Miller indices (h) with (l). In other words, both the conventional cell and reduced cell can be represented by the same experimental d-spacing values reported by Ma and Rubin (2022). In this case, the relative peak intensities ($I_{rel}$) of ($d_{hkl}$) play a role in the selection of the definitive primitive cell from second and third solutions.

The peak intensities depend mainly on the Structure factor (atomic positions in the unit cell) and Atomic scattering factor (atomic number) among other secondary instrumental and textural factors. Usually, and in the case of X-ray data generated from powder X-ray diffractometer, all the intensity factors could be determined using the Rietveld method, implemented in software packages like GSAS (Larson and Von Dreele 2004.) or Fullproff (Carvajal 2001).

In the case of X-ray data provided by Ma and Rubin (2022) that were generated from Electron Backscatter Diffraction (EBSD), the structure was determined by matching both experimental d-spacing and relative intensities simultaneously with those of known structures indexed in X-ray database such as Inorganic Crystal Structure Database (ICSD) and International Centre of Diffraction Data (ICDD).

Given the results reported by the authors, it seems that there was no appropriate procedure used due to the following three details:

a) The cell parameters reported do not resemble those of the primitive cell (see Table 1).
b) While searching in database for matching the experimental d-spacing ($d_{hkl}$) and relative intensities ($I_{rel}$) of known structures, the search procedure should be limited to elements that are mainly in the sample composition; *i.e.*, Fe, Cr and S, and if necessary, another 3d metals. Instead, the authors included the elements Nb and Se in the search procedure.
c) The authors stated that the EBSD patterns of all $FeCr_2S_4$ grains in the Indarch matrix are indexed only by the C2/m $CrNb_2Se_4$ - $Cr_3S_4$ type structure and give a best fit by the synthetic $FeCr_2S_4$ cell from Tressler et al. (1968), in which a = 12.84 Å, b = 3.44 Å, c = 5.94 Å and β = 117°. However, the cell parameters reported by Tressler et al. (1968) are a = 5.94 Å, b = 3.44 Å, c = 11.47 Å and β = 90.85°, which are very similar to those of the reduced cell from Table 1 and resemble the monoclinic I2/m $Cr_3S_4$ type structure (see sec. 2.1).

In the Discussion section of their paper, Ma and Rubin (2022) stated that Zolenskyite ($FeCr_2S_4$) is the Fe-analog of Brezinaite ($Cr_3S_4$), or the Cr-analog of Heideite (ideally $FeTi_2S_4$), in Fig. 2 are shown the experimental X-ray patterns ($d_{hkl}$, $I_{rel}$) of these meteoritic minerals together with synthetic $FeTi_2S_4$ and $Cr_3S_4$ for purpose of comparison, and in Table 2 are shown the corresponding cell parameters (where the third solution from Table 1 - the reduced cell - is proposed for Zolenskyite). As it can be noted, the relative intensities are similar, and also are the cell parameters respectively. Therefore, the proposed structure of Zolenskyite (ideally $FeCr_2S_4$) is the $Cr_3S_4$ type structure, in which Fe atoms are located in $M_I$ layer and Cr atoms are located in $M_{II}$ layer (see Fig. 1), in agreements with previous studies (Bouchard 1967, Tressler et al. 1968, Amiel et al. 2011). In Table 3 are shown the calculated X-ray powder diffraction data for Zolenskyite, provided by Ma and Rubin (2022), and the new Miller indices (also called Bragg refletions) (h,k,l) calculated in this study, which follow the rule h+k+l = 2n (n = integer), characteristic of body-centered I2/m $Cr_3S_4$ type structure (Jellinek 1957).

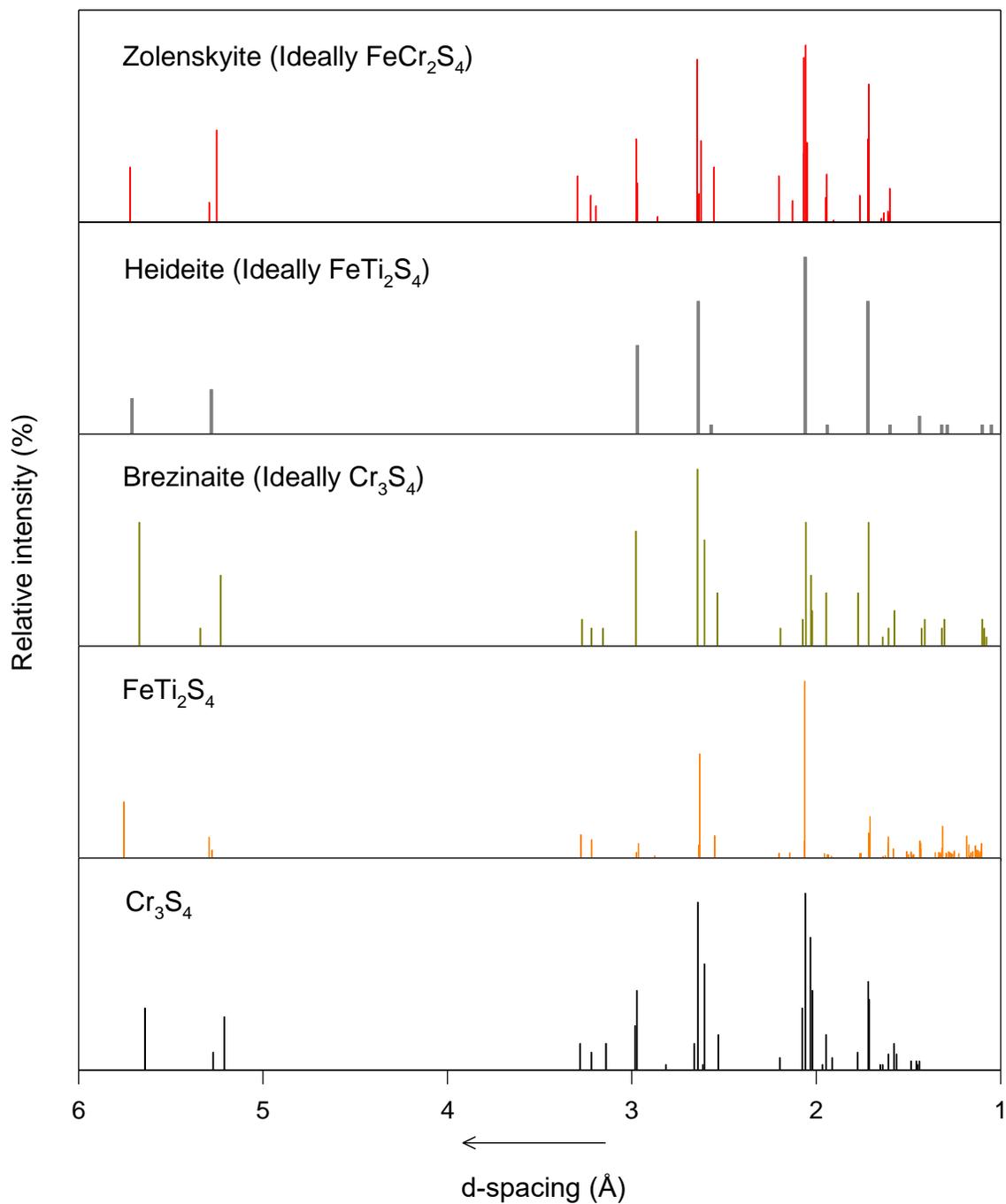

Fig. 2: Experimental X-ray patterns of the meteoritic minerals Zolenskyite (Ma and Rubin 2022), Heideite (Keil and Brett 1974) and Brezinaite (Bunch and Fuchs 1969) and synthetic sulfides $FeTi_2S_4$ (Embaid et al. 2019) and $Cr_3S_4$ (Jellinek 1957).

Table 2: Unit cell parameters of samples with $Cr_3S_4$ type structure.

| Sample | a (Å) | b (Å) | c (Å) | β (°) | Reference |
|---|---|---|---|---|---|
| Zolenskyite | 5.940 | 3.440 | 11.441 | 90.55 | (This study)* |
| Heideite | 5.970 | 3.420 | 11.400 | 90.20 | (Keil and Brett 1974) |
| Brezinaite | 5.960 | 3.425 | 11.270 | 91.53 | (Bunch and Fuchs 1969) |
| $FeTi_2S_4$ | 5.949 | 3.418 | 11.509 | 90.22 | [Embaid et al. 2019] |
| $Cr_3S_4$ | 5.964 | 3.428 | 11.272 | 91.50 | (Jellinek 1957) |
| $FeCr_2S_4$ | 5.940 | 3.440 | 11.470 | 90.85 | [Tressler et al. 1968] |

* Calculated from supplementary X-Ray data provided by Ma and Rubin (2022), see Table 1.

Table 3: Calculated X-ray data for Zolenskyite.

| Ma and Rubin (2022) | | | | | This study | | |
|---|---|---|---|---|---|---|---|
| h | k | l | d [Å] | $I_{rel}$ | h | k | l |
| 2 | 0 | 0 | 5.7203 | 31 | 0 | 0 | 2 |
| 0 | 0 | 1 | 5.2926 | 11 | -1 | 0 | 1 |
| -2 | 0 | 1 | 5.2508 | 52 | 1 | 0 | 1 |
| 1 | 1 | 0 | 3.2943 | 26 | 0 | 1 | 1 |
| 2 | 0 | 1 | 3.2233 | 15 | -1 | 0 | 3 |
| -4 | 0 | 1 | 3.1950 | 9 | 1 | 0 | 3 |
| -1 | 1 | 1 | 2.9768 | 47 | 1 | 1 | 0 |
| -2 | 0 | 2 | 2.9699 | 22 | 2 | 0 | 0 |
| 4 | 0 | 0 | 2.8601 | 3 | 0 | 0 | 4 |
| 0 | 0 | 2 | 2.6463 | 13 | -2 | 0 | 2 |
| 1 | 1 | 1 | 2.6459 | 92 | -1 | 1 | 2 |
| -3 | 1 | 1 | 2.6354 | 16 | 1 | 1 | 2 |
| -4 | 0 | 2 | 2.6254 | 46 | 2 | 0 | 2 |
| 3 | 1 | 0 | 2.5543 | 31 | 0 | 1 | 3 |
| -3 | 1 | 2 | 2.2028 | 26 | 2 | 1 | 1 |
| -6 | 0 | 1 | 2.1283 | 12 | 1 | 0 | 5 |
| 2 | 0 | 2 | 2.0702 | 39 | -2 | 0 | 4 |
| 3 | 1 | 1 | 2.0675 | 93 | -1 | 1 | 4 |
| -5 | 1 | 1 | 2.0575 | 100 | 1 | 1 | 4 |
| -6 | 0 | 2 | 2.0502 | 45 | 2 | 0 | 4 |
| -4 | 0 | 3 | 1.9477 | 14 | 3 | 0 | 1 |
| 1 | 1 | 2 | 1.9428 | 27 | -2 | 1 | 3 |
| 6 | 0 | 0 | 1.9068 | 1 | 0 | 0 | 6 |
| 0 | 0 | 3 | 1.7642 | 15 | -3 | 0 | 3 |
| 0 | 2 | 0 | 1.7200 | 47 | 0 | 2 | 0 |
| -3 | 1 | 3 | 1.7160 | 78 | 3 | 1 | 0 |
| 2 | 2 | 0 | 1.6472 | 2 | 0 | 2 | 2 |
| 0 | 2 | 1 | 1.6358 | 1 | -1 | 2 | 1 |
| -2 | 2 | 1 | 1.6345 | 5 | 1 | 2 | 1 |
| 4 | 0 | 2 | 1.6116 | 6 | -2 | 0 | 6 |
| 3 | 1 | 2 | 1.6095 | 1 | -2 | 1 | 5 |
| 5 | 1 | 1 | 1.6092 | 4 | -1 | 1 | 6 |
| -7 | 1 | 1 | 1.6021 | 19 | 1 | 1 | 6 |

## 3.2. Genesis

Regarding the genesis of Zolenskyite, Ma and Rubin (2022) wrote; "Experiments show that Daubréelite can transform into Zolenskyite at high pressures and moderate temperatures (e.g., 5.5 GPa, 520 $^{o}$C; 3 GPa, 200 $^{o}$C (Tressler et al. 1968). Such conditions likely pertained in highly shocked EH6 chondrites (Rubin and Wasson 2011)".

There are some points that need to be clarified in the above paragraph:

a) There is no report of pressure and temperature sequence such as 3 GPa, 200 $^{o}$C, by Tressler et al. (1968) (see sec. 2.2).
b) Rubin and Wasson (2011) stated that in EH6 chondrites, impacts can produce extreme heterogeneities in the degree of shock damage of target rocks. Shock waves can become chaotic as they interact with inhomogeneities in the rock (e.g., voids, cracks and solid components of different densities). The initial (nanosecond-scale) peak pressure in the shock front and the resulting shock temperature can have grain-to-grain variations of an order of magnitude.
c) The duration of pressure-temperature conditions during shock metamorphism is typically very short, ranging from milliseconds to seconds and this rapid timescale is due to the high-pressure pulses generated by hypervelocity impacts, such as those from meteorites (Hu and Sharp 2022).

Therefore, from the above b and c points we can note that in the geological scale, there are extreme heterogeneities in pressure and temperature values during shock metamorphism and within the time scale of seconds as maximum duration. On the other side and from the laboratory scale, the synthesis procedure of monoclinic $FeCr_2S_4$ is carried out by using static pressures and temperatures during considerable time intervals (from a day to week, see sec. 2.2). DeCarli (2005) draw the attention on our knowledge of shock metamorphism as quite incomplete because of difference between geological scale and laboratory scale conditions.

## 4. OPEN ISSUES

### 4.1. Synthesis vs. genesis

In summary, from the perspective of Solid State laboratory experiments, the monoclinic phase of $FeCr_2S_4$ (ideal composition of Zolenskyite) is obtained through a structural transformation of cubic $FeCr_2S_4$ (ideal composition of Daubréelite) with temperature, pressure, duration and quenching, this fact gives rise to the following question; how to extrapolate this procedure from the laboratory scale to a geological one for the genesis of the meteoritic mineral Zolenskyite? In other words; does the sequence (temperature, pressure, duration) = (1000 $^{o}$C, 6.5 GPa, 1 hour) (Bouchard 1967) or (520 $^{o}$C, 5.5 GPa, 7 days) (Tressler et al. 1968) occur during collisions on the parent body as Ma and Rubin (2022) stated above? And even if these sequences arise naturally, there is another variable for cooling to take into account; fast cooling (quenching) is the condition for the genesis of $FeCr_2S_4$ with monoclinic $Cr_3S_4$ type structure (see sec. 2.2). Otherwise, slow cooling (annealing) causes the formation of $FeCr_2S_4$ with hexagonal NiAs type structure (space group P63/mmc) (Albers and Rooymans 1965).

## 4.2. Ideal and real compositions

There is another issue, that usually is not taken into account in Planetary Science; the real or natural composition of meteoritic minerals – as they are – with minor and trace metals. Ma and Rubin (2022) refer to the natural composition of Zolenskyite $Fe_{0.99}Mn_{0.04}Ca_{0.01}Cr_{1.99}S_{3.98}$ with the ideal formula $FeCr_2S_4$, also the same denomination is used in the literature for Heideite $(Fe, Cr)_{1+x}(Ti, Fe)_2S_4$ - ideally $FeTi_2S_4$ (Keil and Brett 1974, Kurat et al. 2004) - and Brezinaite $(Cr_{2.65}Fe_{0.20}V_{0.09}Ti_{0.06}Mn_{0.04})_{3.04}S_4$ - ideally $Cr_3S_4$ (Bunch and Fuchs 1969) -. These open issues about the Heideite and Brezinaite minerals were discussed in a contemporary review (Embaid 2022) and can be extended for the Zolenskyite, that is; the meteoritic minerals Heideite and Brezinaite are structurally sensitive to the method of synthesis, and so is their magnetic behavior, especially in the presence of Cr atoms in Heideite and Fe atoms in Brezinaite with V, Ti and Mn traces. There is no phase diagram well established for the synthesis of the monoclinic systems leading to the real/natural composition formula of Heideite $(Fe, Cr)_{1+x}(Ti, Fe)_2S_4$ in the Bustee and Kaidun meteorites, nor one of Brezinaite $(Cr_{2.65}Fe_{0.20}V_{0.09}Ti_{0.06}Mn_{0.04})_{3.04}S_4$ in the Tucson meteorite. Indeed, they have never been synthesized as they are, with minor and trace metals.

## 5. IMPLICATIONS

### 5.1. Solid State properties; metallic behavior

Although the monoclinic $FeCr_2S_4$ phase was synthesized and structurally characterized by X-Ray Diffraction a long time ago (Bouchard 1967, Tressler et al. 1968), (see sec. 2.2), no other Solid State characterizations of as the synthesized phase were reported so far. Nevertheless, there is one report about structural, electrical-transport and magnetic properties of spinel $FeCr_2S_4$ phase, studied by X-Ray Diffraction, Electrical Resistance R(T), and $^{57}$Fe Mössbauer Spectroscopy (MS) at variable pressure up to 20 GPa; for pressures above 10 GPa, the spinel $FeCr_2S_4$ transforms to monoclinic $FeCr_2S_4$ type structure, which exhibits a metallic behavior and absence of iron magnetic moment attributed to the Mott transition, as evidenced by R(T) and MS measurements (Amiel et al. 2011). This result encourages further investigation, at first stage, involving the synthesis and study of magnetic properties by MS at variable temperature of monoclinic $FeCr_2S_4$, as done for the synthetic and isostructural sulfides $FeTi_2S_4$ (Embaid et al. 2019) and $FeV_2S_4$ (Embaid and Gonzalez-Jimenez 2013) (see sec. 2.1), whose magnetic properties where analyzed to discuss the possible superconducting property of meteoritic minerals Heideite and Brezinaite, given the natural composition with minor and trace metals (Embaid 2022) (see sec. 4.2).

## 5.2. Artificial origin should not be overlooked; an open debate

Whatever the results to be obtained by addressing the above open issues, they will shed more light on the genesis of Zolenskyite. In this context, it is worth to promote an open debate about the hypothesis of artificial origin, specifically Technosignatures, in the mainstream of the meteoritic minerals Heideite and Brezinaite (Embaid 2022), given that the genesis of these minerals, like Zolenskyite, could require a controlled and sophisticated process not easily found in nature. A comprehensive review about the emerging field of Technosignatures involving the meteoritic minerals Heideite and Brezinaite is published in the after-mentioned reference. However, and for the debate proposal, it is worth to cite textually some quotes and opinions regarding this type of Technosignature search in the literature:

**"Alien artifacts"**
The search for extraterrestrial signals and alien radiation leakage reflects the habit of astronomers to study the emission from celestial bodies. That is why only a negligible part ($3*10^{-9}$) of the Galaxy's lifetime is accessible to the current SETI experiment. However, the search for alien artefact-meteorites accumulated by the Earth could cover the entire history of the Galaxy. - If there are alien artefacts between the stars, some of them are likely to fall to Earth at times - (Arkhipov 1997).

**"Must not overlook"**
We must not overlook the possibility that alien technology has impacted our immediate astronomical environment, even Earth itself, but probably a very long time ago. This raises the question of what traces, if anything, might remain today (Davies 2012).

**"Non-terrestrial artifacts (NTAs)"**
Our intention is to present a framework by which we can estimate the completeness of our search for NTAs, which will inevitably increase as we continue to explore the moon, Mars, and other nearby regions of space. The discovery of extraterrestrial technology would certainly be one of the most significant findings in human history; for even if this technology were non-functional, it would give us some certainty that life - and intelligence - has developed elsewhere. With so many places for a such small observational probe to hide in our own backyard, we may as well keep our eyes open (Haqq-Misra and Kopparapu 2012).

**"Drake equation for alien artifacts"**
I propose a version of the Drake equation to include searching for alien artifacts, which may be located on the Moon, Earth Trojans, and Earth co-orbital objects. The virtue of searching for artifacts is their lingering endurance in space, long after they go dead. I compare a search for extraterrestrial artifacts (SETA) strategy with the existing listening to stars search for extraterrestrial intelligence (SETI) strategy (Benford 2021).

**"Derelict technology"**
Artifacts in the Solar System might be active, producing communicative signals, generating heat, maintaining attitude, and maneuvering; or they might be long dead, subject to collisions with asteroids or meteorites, or lying beneath the regolith or sands of a terrestrial body (Wright, (2021).

**"Search for extraterrestrial artifacts"**
In this article, we suggest that the systematic exclusion of the search for extraterrestrial artifacts from mainstream research, despite the technological feasibility of this search and overwhelming public interest in it, reveals an institutional bias against novel research paradigms, indicating it is time for astronomers to embrace a bolder scientific program (Eldadi et al. 2025).

## ACKNOWLEDGMENTS

The author is grateful to the colleague Jacob Haqq-Misra (Blue Marble Space Institute of Science, Seattle, WA, USA) for his fruitful discussions and comments that helped shape this manuscript.